\documentclass[twocolumn,showpacs,aps,prl]{revtex4}
\usepackage{amsmath,amsfonts,bm,graphicx}

\newcommand{\G}{\Gamma}

\newcommand{\eps}{\epsilon}
\renewcommand{\t}{\tau}
\newcommand{\s}{\sigma}
\renewcommand{\t}{\tau}

\newcommand{\phd}{\phantom{\dagger}}

\begin{document}

\title{Symmetry fingerprints of a benzene single-electron
transistor}

\author{Georg Begemann}

\affiliation{Theoretische Physik, Universit\"{a}t Regensburg,
93040 Regensburg, Germany}

\author{Dana Darau}

\affiliation{Theoretische Physik, Universit\"{a}t Regensburg,
93040 Regensburg, Germany}

\author{Andrea Donarini}

\affiliation{Theoretische Physik, Universit\"{a}t Regensburg,
93040 Regensburg, Germany}

\author{Milena Grifoni}

\affiliation{Theoretische Physik, Universit\"{a}t Regensburg,
93040 Regensburg, Germany}

\date{\today}

\begin{abstract}

The interplay between Coulomb interaction and orbital symmetry
produces specific transport characteristics in molecular single
electron transistors (SET) that can be considered as the
fingerprints of the contacted molecule. Specifically we predict,
for a benzene SET, selective conductance suppression and the
appearance of negative differential conductance when changing the
contacts from para to meta configuration. Both effects originate
from destructive interference in transport involving states with
\emph{orbital degeneracy}.

\end{abstract}


\maketitle

Understanding the conduction characteristics through single
molecules is one of the crucial issues in molecular electronics
\cite{Cuniberti-book}. The dynamics of the electron transfer to
and from the molecule depends on the intrinsic electronic spectrum
of the molecule as well as on the electronic coupling of the
molecule to its surroundings.

In recent years the measurement of stability diagrams of single
electron transistor (SET) devices has become a very powerful tool
to do spectroscopy of small conducting systems via transport
experiments. Thus the capability to perform  three terminal
measurements on single molecules
\cite{ParkJ02,Liang02,Kubatkin03,Yu05,Danilov08,Chae06,Poot06,Heersche06,Osorio07}
has been a fundamental achievement for molecular electronics. Such
molecular transistors might display transport properties which are
very different from those of conventional SETs.  In fact,
vibrational or torsional modes \cite{Chae06,Osorio07} and
intrinsic symmetries/asymmetries of the molecule can hinder or
favor transport through the SET, visible e.g. in the
absence/presence of specific excitation lines in the stability
diagram of the molecular SET, or in negative differential
conductance features. Many-body phenomena as e.g. the Kondo
effect, have been observed as well
\cite{ParkJ02,Liang02,Yu05,Osorio07}.

Despite the experimental progress, the theoretical understanding
of the properties of single organic molecules coupled to
electrodes is far from being satisfactory. On the one hand,
numerical approaches to transport based e.g. on the combination of
Green's function methods with density functional theory have
become a standard approach to study transport at the nanoscale
\cite{Cuniberti-book}.
However, this technique is not appropriate for the description of
transport through a molecule weakly coupled to leads, due to the
crucial role played by the Coulomb interaction in these systems.
Hence, in \cite{Hettler03}, an electronic structure calculation
for a benzene molecule was performed in order to arrive at an
effective interacting Hamiltonian for the $\pi$ orbitals, to be
solved to determine the I-V characteristics of a benzene junction.
\begin{figure}[h!]
  \includegraphics[width=0.9\columnwidth,angle=0]{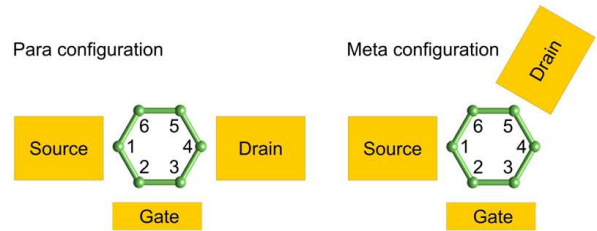}
  \caption{(color online) The two different setups for the benzene SET considered in this letter.
  }
  \label{fig1}
\end{figure}
In this letter we consider the electronic transport through a
benzene SET. Similar to \cite{Hettler03}, in order to devise a
semi-quantitative description, we start from an interacting
Hamiltonian of isolated benzene where only the localized $p_z$
orbitals are considered and the ions are assumed to have the same
spatial symmetry as the relevant electrons. The Hamiltonian for
the isolated molecule possesses $4^6=4096$ eigenstates, to be
calculated numerically, and whose symmetries can be established
with the help of group theory.  Large
degeneracies of the electronic states occur. For example, while
the six-particles ground state ($A_{1g}$ symmetry) is
non-degenerate, there exist four seven-particle ground states due
to spin and orbital ($E_{2u}$ symmetry) degeneracy.  When coupling
the benzene SET to leads in the meta and para configurations,
these orbitals symmetries lead to very different stability
diagrams for the two configurations (see Fig.~\ref{fig2}).
Striking are the selective reduction of conductance
(Fig.~\ref{fig3}) and the occurrence of negative differential
conductance (NDC) features when changing from para to
meta-configurations. As shown in Fig.~\ref{fig4}, the NDC effect
occurs due to the formation of a blocking state at certain values
of the bias voltage. The blocking is clearly visible by monitoring
the position-dependent many-body transition probabilities which,
at given values of the bias voltage, can exhibit  nodes at the
same position as one of the contacts. NDC for benzene junctions
has been predicted also in \cite{Hettler03}, but in the para
configuration \emph{and} in presence of an external
electromagnetic field. In our work NDC occurs despite  the absence
of an external field. Both the effects we predict originate from
bias dependent interference of orbitally degenerate states:
coherences, neglected in \cite{Hettler03}, are essential to
capture interference effects when solving the equations for the
benzene's occupation probabilities. Interference phenomena in
transport through benzene have been recently discussed also in
\cite{Cardamone06, Gagliardi07}. The parameter regime is however
very different, as both discuss the strong tunneling limit, where
Coulomb blockade effects are not relevant.

We start from the total Hamiltonian
 $H = H_{\rm ben} + H_{\rm leads} + H_{\rm T}$,
where the Hamiltonian for benzene reads:
\begin{equation}
    \begin{split}
    H_{\rm ben} = \,\,
    &\xi_0 \sum_{i\s} d^{\dagger}_{i\s}d^{\phd}_{i\s} +
    b  \sum_{i\s}\left(d^{\dagger}_{i\s}d^{\phd}_{i+1\s} + d^{\dagger}_{i+1\s}d^{\phd}_{i\s}\right)\\
    +& U \sum_i \left(n_{i\uparrow} - \tfrac{1}{2}\right)
                \left(n_{i\downarrow} - \tfrac{1}{2}\right)\\
    +& V \sum_i\left(n_{i\uparrow} + n_{i\downarrow}- 1\right)
               \left(n_{i+1\uparrow} + n_{i+1\downarrow} -
               1\right).
    \end{split}
    \label{eq:PPP}
\end{equation}
Here $d^{\dagger}_{i\sigma}$ creates an electron of spin $\sigma$
in the $p_z$ orbital of carbon $i$, $i = 1,\ldots,6$ runs over the
six carbon atoms of benzene and $n_{i\sigma} =
d^{\dagger}_{i\sigma} d^{\phd}_{i\sigma}$. This Hamiltonian is
respecting the $D_{6h}$ symmetry of benzene and also the
particle-hole symmetry. Mechanical oscillations are at this level
neglected and all the atoms are considered in their equilibrium
position. The parameters $b$, $U$, and $V$ for isolated benzene
are given in the literature \cite{Barford-book} and  are chosen to
fit excitation spectra. Even if the presence of metallic
electrodes is expected to cause a substantial renormalization of
$U$ and $V$, we do not expect the main results of this work to be
affected by this change. The weak coupling suggests that the
symmetry of the molecule will remain unchanged and with it the
structure of the Hamiltonian \eqref{eq:PPP}. The gate voltage
$V_g$ is introduced by a renormalized on-site energy $\xi = \xi_0
- eV_g$ and we conventionally set $V_g = 0$ at the charge
neutrality point. We represent source and drain leads  as two
resevoirs of non- interacting electrons: $H_{\rm leads} =
\sum_{\alpha\,k\,\s}(\eps_k - \mu_{\alpha})
 c^{\dagger}_{\alpha k \s}c^{\phd}_{\alpha k \s}$,
where $\alpha = L,\,R$ and the chemical potentials $\mu_{\alpha}$
of the leads depend on the applied bias voltage $\mu_{L,R} = \mu_0
\pm \tfrac{V_b}{2}$. In the following we will measure the energy
starting from the equilibrium chemical potential $\mu_0 = 0$ thus
giving a negative energy to the holes in equilibrium. The coupling
to source and drain leads is described by
%
\begin{equation}
 H_{\rm T} = t\sum_{\alpha k \s}
 \left(d^{\dagger}_{\alpha \s} c^{\phd}_{\alpha k \s} +
       c^{\dagger}_{\alpha k \s} d^{\phd}_{\alpha \s}\right),
\end{equation}
where we define $d^{\dagger}_{\alpha \sigma}$ as the creator of
the electron in the benzene carbon atom  which is closer to the
lead $\alpha$. In particular  $d^{\dagger}_{R \sigma} :=
d^{\dagger}_{4 \sigma}, d^{\dagger}_{5 \sigma}$ respectively in
the para and meta configurations, while $d^{\dagger}_{L \sigma} :=
d^{\dagger}_{1 \sigma}$ in both setups. Due to the weak coupling
to the leads we can assume that the potential drop is all
concentrated at the lead-molecule interface and is not affecting
the molecule itself.

Given the high degeneracy of the spectrum, the method of choice to
treat the dynamics in the weak coupling is the Liouville equation
method already used e.g. in \cite{Donarini06, Mayrhofer07}.
Starting point is the Liouville equation for the reduced density
operator $\dot{\sigma} ={\rm Tr}_{\rm leads}\{\dot{\rho}\} =
-\frac{i}{\hbar}{\rm Tr}_{\rm leads}\{[H,\rho]\}$ where $\rho$ is
the density operator \cite{Blum-book}. Due to the weak coupling to
the leads we treat the effects of $H_{\rm T}$ to the lowest non-
vanishing order.
The reduced density operator $\sigma$ is defined on the Fock space
of benzene but coherences between states with different particle
number and different energy can be neglected, the former because
decoupled from the dynamics of the populations, the latter being
irrelevant due to their fast fluctuation (secular approximation).
As a result we arrive at a generalized master equation (GME) where
coherences between degenerate states are retained. This approach
is robust against the small asymmetries introduced in the molecule
by the coupling to the leads or by deformation as far as the
energy splitting that lifts the orbital degeneracy is comparable
to the thermal energy. The GME is conveniently expressed in terms
of the reduced density operator $\s^{NE} = \mathcal{P}_{NE} \,\s\,
\mathcal{P}_{NE}$, where $\mathcal{P}_{NE} := \sum_{\ell \t} |N\,
E\, \ell\, \t \rangle \! \langle N\, E\, \ell\, \t |$ is the
projection operator on the subspace of $N$ particles and energy
$E$. The sum runs over the orbital and spin quantum numbers $\ell$
and $\tau$, respectively. Eventually the GME reads
\begin{widetext}
\begin{equation}
    \begin{split}
    \dot{\s}^{NE}
    =& -\sum_{\alpha\t} \frac{\G_\alpha}{2}
    \Bigg\{
    d^{\phd}_{\alpha\t}
    \left[f^+_\alpha(H_{\rm ben} - E) + \frac{i}{\pi}\,p_{\alpha}(H_{\rm ben} - E)\right]
    d^{\dagger}_{\alpha\t}\,\s^{NE} +\\
    &\phantom{-\sum_{\alpha\t}\frac{\G_\alpha}{2}\,\,\,}
  + d^{\dagger}_{\alpha\t}
    \left[f^-_\alpha(E - H_{\rm ben}) - \frac{i}{\pi}\,p_{\alpha}(E - H_{\rm ben})\right]
    d^{\phd}_{\alpha\t}\,\s^{NE} + h.c.
    \Bigg\} + \\
& + \sum_{\alpha\t E'} \G_\alpha \mathcal{P}_{NE}
    \Bigg\{
    d^{\dagger}_{\alpha\t} f^+_\alpha(E - E')\, \s^{N-1 E'} d^{\phd}_{\alpha\t}
  + d^{\phd}_{\alpha\t} f^-_\alpha(E' - E)\,\s^{N+1 E'}d^{\dagger}_{\alpha\t}
    \Bigg\}
    \mathcal{P}_{NE},
    \end{split}
\label{eq:GME}
\end{equation}
\end{widetext}
where $\Gamma_{L,R}=
\tfrac{2\pi}{\hbar}|t_{L,R}|^2\mathcal{D}_{L,R}$ equal the bare
transfer rates  with the constant densities of states of the leads
$\mathcal{D}_{L,R}$. Terms describing sequential tunnelling from
and to the lead $\alpha$ are proportional to the Fermi function
$f(x - \mu_\alpha):=f^+_\alpha(x)$ and $f^-_\alpha(x) = 1 -
f^+_\alpha(x)$, respectively. Still in the sequential tunnelling
limit, but due to the presence of coherences, also energy
non-conserving terms are appearing in the generalized master
equation, they are proportional to the function $p_{\alpha}(x) =
-{\rm Re}\psi\left[\tfrac{1}{2} + \tfrac{i\beta}{2\pi}(x -
\mu_\alpha)\right]$ where $\psi$ is the digamma function
\cite{Mayrhofer07,Blum-book}.
\begin{figure}[h!]
  \includegraphics[width=0.9\columnwidth,angle=0]{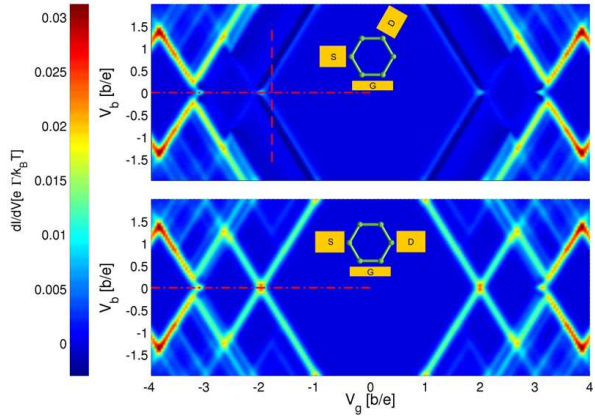}
  \caption{(color online) Stability diagram for the benzene SET
  connected in the meta (above) and para (below) configuration.
  Red dot-dashed lines highlight the conductance cuts
  presented in Fig.~\ref{fig3}, the red dashed line the region corresponding
  to the current trace presented in Fig.~\ref{fig4}. The
  parameters used are $U=4|b|,\,V = 2.4|b|,\,T = 0.04|b|,\,\Gamma =
10^{-3}|b|$.
  }
  \label{fig2}
\end{figure}
Finally we write the GME in the basis of the energy eigenstates
for isolated benzene and find numerically the stationary solution.

A closer analysis of the master equation allows also to define a
current operator (one per molecule-lead contact)
\begin{equation}
    \begin{split}
    \hat{I}_\alpha = \sum_{NE\tau} \mathcal{P}_{NE}
    \Big[ &
    d^{\phd}_{\alpha\tau}f^+_\alpha(H_{\rm PPP}-E)d^{\dagger}_{\alpha\tau}
    +\\
    &-d^{\dagger}_{\alpha\tau}f^-_\alpha(E-H_{\rm PPP})d^{\phd}_{\alpha\tau}
    \Big] \mathcal{P}_{NE}
    \end{split}
\end{equation}
and calculate the stationary current as the average  $I_{\rm L} =
{\rm Tr}\{{\sigma}_{\rm stat} \hat{I}_{\rm L}\} = -I_R$, with
${\sigma}_{\rm stat}$ the stationary density operator.
In Fig.~\ref{fig2} we present the stability diagram for the
benzene SET contacted in the para (lower panel) and meta position
(upper panel). Bright ground state transition lines delimit
diamonds of zero differential conductance typical of the Coulomb
blockade regime while a rich pattern of satellite lines represents
the transitions between excited states. Though several differences
can be noticed, most striking are the suppression of conductance
and the appearance of NDC when passing from para to meta
configuration.
\begin{figure}[h!]
\vspace{5mm}
  \includegraphics[width=0.9\columnwidth,angle=0]{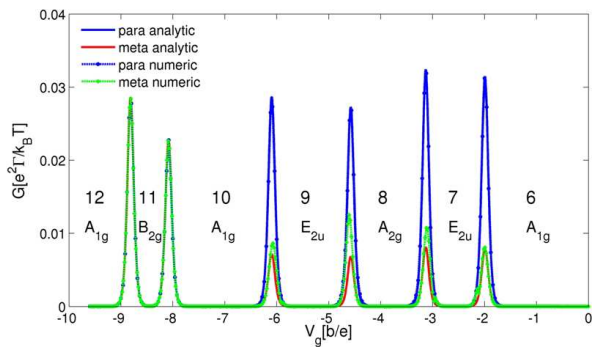}
  \caption{(color online) Conductance of the benzene SET as a function of the gate voltage.
  Clearly visible are the peaks corresponding to the transitions between ground states with
  $N$ and $N+1$ particles. In the low conductance valleys the state of the system
  has a definite number of particles and symmetry as indicated. Selective conductance suppression
  when changing from the meta to the para configuration is observed.}
  \label{fig3}
\vspace{-10mm}
\end{figure}
A zero bias cut of the stability diagrams as a function of the
gate voltage $V_g$ is plotted in Fig.~\ref{fig3}. Only transitions
between ground states are relevant for the conductance. The number
of $p_z$ electrons on the molecule and the symmetry of the ground
state corresponding to the conductance valleys are reported. The
conductance in the meta and para configuration is the same for the
$N=11 \leftrightarrow 12$ and $N=10 \leftrightarrow 11$
transitions while it is systematically suppressed in all other
cases. In other terms transitions between states with $A$ or $B$
symmetry, which do not have orbital degeneracy,  are invariant
under configuration change; transitions that involve an $E$
symmetry, and hence imply twofold orbital degeneracy, are
suppressed. \emph{Destructive interference} between orbitally
degenerate states explains the systematic conductance suppression.
By neglecting the energy non-conserving terms in \eqref{eq:GME} we
derived an analytical formula for the conductance close to the
resonance between $N$ and $N+1$ particle states:
\begin{widetext}
\begin{equation}
G_{N,N+1}(\Delta E) = 2e^2\frac{\Gamma_L \Gamma_R}{\Gamma_L +
\Gamma_R}
 \frac{\left|
 \sum_{nm\tau}
 \langle N,\,n|d^{\phd}_{L\tau}|N+1,\,m\rangle
 \!
 \langle N+1,\,m |d^{\dagger}_{R\tau}|N,\,n\rangle
 \right|^2}
 {\sum_{nm\alpha\tau}|\langle
 N,\,n|d^{\phd}_{\alpha\tau}|N+1,\,m \rangle|^2}
 \left[-\frac{f'(\Delta E)}{(S_{N+1}-S_N)f(\Delta E)+S_N}\right]
\end{equation}
\end{widetext}
where $\Delta E = E_{g,N} - E_{g,N+1} + e V_g$ is the energy
difference between the benzene ground states with $N$ and $N+1$
electrons diminished by a term linear in the side gate, $n$ and
$m$ label the $S_N$-fold and $S_{N+1}$-fold degenerate ground
states with $N$ and $N+1$ particles, respectively. Interference
effects are contained in the numerator of the third factor
(overlap factor $\Lambda$). In order to make these more visible we
remind that $d^{\dagger}_{R\tau} = \mathcal{R}^{\dagger}_\phi
d^{\dagger}_{L\tau} \mathcal{R}_\phi$, where $\mathcal{R}_\phi$ is
the rotation operator of an angle $\phi$ and $\phi = \pi$ for the
para while $\phi = 2\pi/3$ for the meta configuration. All
eigenstates of $H_{\rm ben}$ are eigenstates of the discrete
rotation operators with angles multiples of $\pi/3$ and the
eigenvalues are phase factors. The overlap factor now reads:
\begin{equation}
\Lambda =
 \left|
 \sum_{nm\tau}
 |\langle N,\,n|d^{\phd}_{L\tau}|N+1,\,m\rangle|^2 e^{i\phi_{nm}}
 \right|^2,
\end{equation}
where $\phi_{nm}$ encloses the phase factors coming from the
rotation of the states $|N,n\rangle$ and $|N+1,m\rangle$.
Interference is possible only when $S_N$ or $S_{N+1} > 1$, that is
in presence of degenerate states. It generates a considerable
reduction by passing from the para to the meta configuration as
seen in Fig. \ref{fig3}.
\begin{figure}[h!]
  \includegraphics[width=0.9\columnwidth,angle=0]{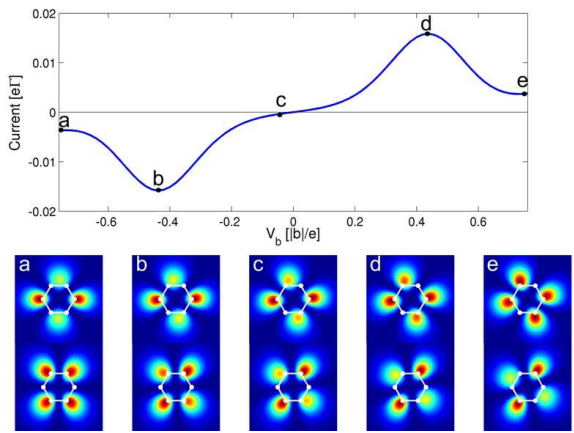}
  \caption{(color online) Upper panel - Current through the benzene SET
  in the meta configuration calculated at  bias and gate voltage conditions indicated by
  the dashed line of Fig. \ref{fig2}. A pronounced NDC is visible.
  Lower panels - Transition probabilities between the 6 particle and the
  7 particle ground states for bias voltage values labelled $a-e$ in the upper panel.
  The transitions to a blocking state is visible in the upper (lower)
  part of the $e$ ($a$) panels.}
  \label{fig4}
\end{figure}
Interference also affects non- linear transport and produces in
the meta configuration NDC at the border of the 6 particles state
diamond (Fig. \ref{fig2}). The upper panel of Fig. \ref{fig4}
shows the current through the benzene SET contacted in the meta
configuration as a function of the bias voltage. The current is
given for parameters corresponding to the red dashed line of Fig.
\ref{fig2}. In this region only the 6 and 7 particles ground
states are populated. The 6 particle ground state is not
degenerate. The 7 particle ground state is 4-fold degenerate,
though the twofold spin degeneracy is not important since spin
coherences vanish in the stationary limit and the $S_z = 1/2$ and
$-1/2$ density matrices are equal for symmetry. At low bias the 6
particle state is mainly occupied. As the bias is raised
transitions $6 \leftrightarrow 7$ occur and current flows. Above a
certain bias threshold a blocking state is populated and the
current is reduced. To visualize this, we introduce the
probability (averaged over the $z$ coordinate)
\begin{equation}
P(x,y; \ell) = \lim_{L \to \infty} \sum_\t\frac{1}{L}
\int_{-L/2}^{L/2} {\rm d}z |\langle 7g \ell \t |
\psi^{\dagger}(\vec{r})| 6g \rangle|^2
\end{equation}
for benzene to make a transition between the state $| 6g \rangle$
and one of the states $| 7g \ell \tau \rangle$ by adding or
removing an electron in position $\vec{r}$. Each of the lower
panels of Fig. \ref{fig4} are surface plots of $P(x,y; \ell)$ for
the 7 particles basis that diagonalizes the stationary density
matrix at a fixed bias. The upper plot of the $e$ panel describes
the transitions to the blocking 7 particle state that accepts
electrons from the source lead (close to the carbon 1) but cannot
release electrons to the drain (close to carbon 5). The energy
non-conserving rates prevent the complete efficiency of the
blocking by ensuring a slow depopulation of the blocking state.
At large negative bias the blocking scenario is depicted in the
panel $a$. We remark that only a description that retains
coherences between the degenerate 7 particle ground states
correctly captures NDC at both positive and negative bias.

To summarize, we analyzed the transport characteristics of a
benzene based SET. The interplay between Coulomb interaction and
orbital symmetry is manifested in a destructive interference
involving orbitally degenerate states, leading to selective
conductance suppression and negative differential conductance when
changing the contacts from para to meta configuration.

We acknowledge financial support by the DFG within the research
programs SPP 1243 and SFB 689.

\end{document}